\documentclass[aip,apl,amsmath,amssymb,superscriptaddress,reprint]{revtex4-2}

\usepackage{graphicx}
\usepackage{dcolumn}
\usepackage{bm}

\usepackage[utf8]{inputenc}
\usepackage[T1]{fontenc}
\usepackage{mathptmx}

\usepackage{gensymb}
\usepackage{upgreek}
\usepackage{todonotes}

\usepackage{xr}
\begin{document}

\title{Phase domain boundary motion and memristance in gradient-doped FeRh nanopillars induced by spin injection}

\author{Rowan~C.~Temple}
\affiliation{School of Physics and Astronomy, University of Leeds, Leeds LS2 9JT, UK}

\author{Mark~C.~Rosamond}
\affiliation{School of Electronic and Electrical Engineering, University of Leeds, Leeds LS2 9JT, UK}

\author{Jamie~R.~Massey}
\affiliation{School of Physics and Astronomy, University of Leeds, Leeds LS2 9JT, UK}

\author{Trevor~P.~Almeida}
\affiliation{SUPA, School of Physics and Astronomy, University of Glasgow, Glasgow G12 8QQ, UK}

\author{Edmund~H.~Linfield}
\affiliation{School of Electronic and Electrical Engineering, University of Leeds, Leeds LS2 9JT, UK}

\author{Damien~McGrouther}
\affiliation{SUPA, School of Physics and Astronomy, University of Glasgow, Glasgow G12 8QQ, UK}

\author{Stephen~McVitie}
\affiliation{SUPA, School of Physics and Astronomy, University of Glasgow, Glasgow G12 8QQ, UK}

\author{Thomas~A.~Moore}
\affiliation{School of Physics and Astronomy, University of Leeds, Leeds LS2 9JT, UK}

\author{Christopher~H.~Marrows}
\email{c.h.marrows@leeds.ac.uk}
\affiliation{School of Physics and Astronomy, University of Leeds, Leeds LS2 9JT, UK}

\date{\today}

\begin{abstract}

The B2-ordered alloy FeRh shows a metamagnetic phase transition, transforming from antiferromagnetic (AF) to ferromagnetic (FM) order at a temperature $T_\mathrm{t} \sim 380 $~K in bulk. As well as temperature, the phase transition can be triggered by many means such as strain, chemical doping, or magnetic or electric fields. Its first-order nature means that phase coexistence is possible. Here we show that a phase boundary in a 300~nm diameter nanopillar, controlled by a doping gradient during film growth, is moved by an electrical current in the direction of electron flow. We attribute this to spin injection from one magnetically ordered phase region into the other driving the phase transition in a region just next to the phase boundary. The associated change in resistance of the nanopillar shows memristive properties, suggesting potential applications as memory cells or artificial synapses in neuromorphic computing schemes.

\end{abstract}

\maketitle

Ever since the discovery of the large resistivity drop \cite{Kouvel1962} in B2-ordered FeRh at its phase transition into the FM state, electrical currents have been used as a probe of the phase state \cite{Algarabel1995,Sharma2011,DeVries2013,Uhlir2016,Matsumoto2018}. Device proposals based on the electrical properties of the phase transition include driving it with an electric field \cite{Cherifi2014,Suzuki2014} or an AF memory resistor that is written in the FM state \cite{Marti2014,Moriyama2015}.

Nevertheless, the use of electrical currents to drive, rather than simply probe, the phase transition has received less attention \cite{Lewis2016}. A trivial example is to drive the transition by Joule heating \cite{Matsuzaki2015,Moriyama2015}. Of more interest is the direct influence of the current on the transition by means of spintronic effects where electrically injected spins \cite{vanSon1987} may drive the transition by favouring the FM phase \cite{Zyuzin2010}. A tentative early observation was that an FeRh wire showed a current-induced phase transition at a lower current density when the current passed through overlaid spin-polarised Co wires rather than unpolarised Cu wires \cite{Naito2011}. The effect was more clearly seen when current was injected through a Co/FeRh interface, showing a marked suppression of the AF phase for current densities on the scale of $10^{11}$~A/m$^2$, which was not present when a Cu/FeRh interface was used \cite{Suzuki2015}.

Here we show that passing current through an AF/FM phase boundary within an FeRh nanopillar reversibly drives the AF$\leftrightarrow$FM phase transition, realising a decade-old prediction of a self-propelled interface driven by spin injection into a metamagnet \cite{Zyuzin2010}. Combining the ideas of the current both driving and detecting the phase transition means that our nanopillars show memristive behaviour. Originally proposed to complete the set of passive circuit elements \cite{Chua1971}, memristors were first realised in a titanium dioxide nanopillar in which current-driven ionic transport shifts the boundary between an undoped TiO$_2$ layer and a doped TiO$_{2-x}$ region \cite{Strukov2008}. There is an analogy between the motion of the oxidation front there and the AF/FM phase boundary in our metamagnetic nanopillars.

\begin{figure*}[tb]
\includegraphics[width=16cm]{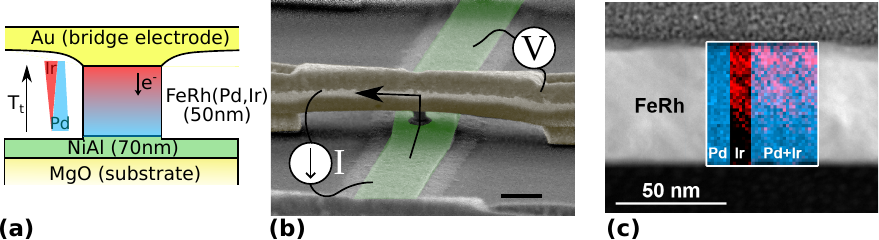}
\caption{Nanopillar design and fabrication. (a) Schematic diagram of a nanopillar with Pd to Ir doping gradient indicated. The electron flow shown is for positive current. (b) Scanning electron micrograph of a 300~nm diameter doping-gradient FeRh nanopillar with flying bridge contact. Current source and voltmeter connections are indicated. The scale bar is 500~nm. (c) High angle annular dark field image of cross-section of gradient-doped FeRh continuous film with superimposed energy dispersive X-ray spectroscopy (EDS) data showing the Pd/Ir concentration gradient.}
\label{nanopillar}
\end{figure*}

Fig.~\ref{nanopillar}a schematically illustrates the nanopillars that we have studied. They were patterned from gradient-doped FeRh epilayers grown on MgO substrates with a NiAl buffer layer to ensure epitaxial growth, from which a bottom contact is formed. A top contact is made using a polycrystalline Au flying bridge. A scanning electron micrograph of a completed device is shown in Fig.~\ref{nanopillar}b. The dopants are Pd, which reduces $T_\mathrm{t}$, and Ir, which causes $T_\mathrm{t}$ to rise\cite{Kouvel1966,Barua2013b}, arranged with a gradient so that the epilayer is Pd-rich close to the NiAl buffer and Ir-rich near to its top surface, leading to a gradient in $T_\mathrm{t}$ through the nanopillar height\cite{LeGraet2015}. Further details of the growth and patterning methods are given in the Supplementary Information. Fig.~\ref{nanopillar}c shows a transmission electron microscopy cross-section from an unpatterned gradient-doped FeRh film grown in the same way, together with an elemental map confirming the doping gradient. The consequence of this gradient is that for a wide range of temperatures, a height within the nanopillar can be found where the temperature corresponds to the local $T_\mathrm{t}$, and a horizontal phase boundary then separates the nanopillar into a FM region below that height and an AF region above it.

When an electrical current flows vertically through the nanopillar it must pass through this horizontal magnetic phase boundary. Our nanopillar device is connected so that a positive flow of conventional current is from bottom to top (see Fig.~\ref{nanopillar}a), meaning that electrons flow from the AF region to the FM region. In this case, electrons with no net spin-polarisation are driven into the FM region. On the other hand, when a negative conventional current flows, spin-polarised electrons are driven from the FM into the AF region, a phenomenon known as electrical spin injection, where a small non-equilibrium magnetisation is generated close to the interface in the non-ferromagnetic material\cite{vanSon1987}. This effect--and its inverse, when the ferromagnet is slightly depolarised by current flow in the opposite direction--is expected to trigger the phase transition near to the phase boundary \cite{Zyuzin2010}. This causes motion of the phase boundary in a direction that can be selected by the direction of current flow. Since the AF phase is more resistive than the FM phase\cite{Kouvel1966}, this motion changes the series resistance of the nanopillar, providing our means for detecting the effect.

\begin{figure}[tb]
\includegraphics[width=7.5cm]{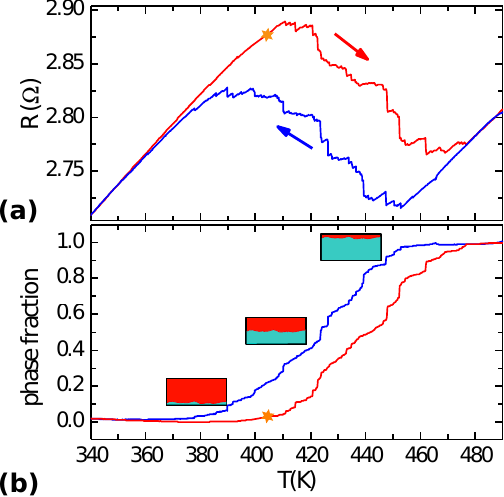}
\caption{Phase transition detected by resistance measurement. (a) Resistance of a 300~nm diameter  nanopillar as changing temperature induces the phase transition. Heating and cooling branches are marked in the red and blue, respectively. A small test current of 300~$\mu$A was used. (b) Applying a linear transformation based on series resistors to the resistance gives the phase fraction that has transformed to become ferromagnetic at each temperature. The inset diagrams indicate the approximate phase state at selected temperatures, with red showing the AF phase and green the FM phase. The star on the heating branch indicates the temperature for the current pulse measurements in Fig.~\ref{pulsemem}.}
\label{transition}
\end{figure}

The nanopillar resistance $R$ was measured using the quasi-four point method illustrated in Fig.~\ref{nanopillar}b, with further details given in the supplementary information. The variation of resistance with temperature $T$ for a 300~nm diameter nanopillar is shown in Fig.~\ref{transition}a. The usual hysteresis for a first-order phase transition is evident. The transition is very broad, spanning a range from about 370-475~K. The limits of this temperature range represent the extreme values of $T_\mathrm{t}$ at the most Pd-rich and Ir-rich points at the bottom and top of the nanopillar. Below 370~K, the entire nanopillar is in the AF phase, whilst above 475~K the FM phase occupies the whole structure. As an aside, it is worth noting that the fact that the phase boundary moves vertically in our pillar, owing to the doping-induced transition temperature gradient, means that our pillars do not show the few large jumps on the cooling branch that have been seen in other nanoscale FeRh structures \cite{Uhlir2016,Matsumoto2018}.

Within this range of temperatures, the nanopillar is divided into two regions in different phases. As the temperature rises and falls a horizontal phase boundary sweeps up and down the nanopillar, as shown schematically in the insets in Fig.~\ref{transition}b. We can gain a clear view of how the phase fraction (or equivalently the position in height of the phase boundary) varies as the temperature changes by noting that the measured resistance is simply the series resistance of the two phase regions (Fig.~\ref{transition}b). Superimposed on the resistance change caused by the phase transformation is the usual linear rise in the resistance of a metal as its temperature increases, which is subtracted out. The resistance does not vary smoothly with temperature, showing many abrupt jumps and steps that represent the phase boundary jumping between pinning sites as it travels the height of the nanopillar\cite{Uhlir2016}.

The measurements shown in Fig.~\ref{transition} were performed using a current of only 300~$\upmu$A. This corresponds to a current density of about $4 \times 10^9$~A/m$^2$, which is too small to noticeably affect the transition. (Current densities are estimated from the patterned pillar diameter, neglecting the smaller contact area from the bridge and any current-crowding effects.) To attempt to do so, we applied pulse trains of $10^6$ pulses, each of 1~$\upmu$s duration, with a duty cycle of 10\% to the 300~nm nanopillar. The amplitude of each pulse was 20~mA, corresponding to a current density of $\sim 3 \times 10^{11}$~A/m$^2$. Control measurements at 320~K, outside the transition region, showed small rises in resistance $\Delta R \approx 3$~m$\Omega$, regardless of the current direction, consistent with Joule heating. Further details are given in the supplementary information.

In Fig.~\ref{pulsemem} we show the effect of similar pulse trains at 404~K on the heating branch of the phase transition hysteresis loop, indicated by the gold star in Fig.~\ref{transition}a. At this temperature there will be a small region of FM phase occupying the bottom of the nanopillar, separated from the rest of the nanopillar by a phase boundary. In this case there are three important differences in the response with respect to the control measurements. The resistance change is negative and also much larger than at 320~K, on the scale of a few tens of m$\Omega$. These two differences confirm that the nanopillar has indeed entered the mixed phase regime, where heating causes an increase in the FM phase fraction and a concomitant drop in resistance. Each excursion in $\Delta R$ lasts for a few seconds and ends abruptly, which we associate with the sudden motion of the phase boundary between pinning sites as the substrate cools after the pulse.

\begin{figure}[tb]
\includegraphics[width=7.5cm]{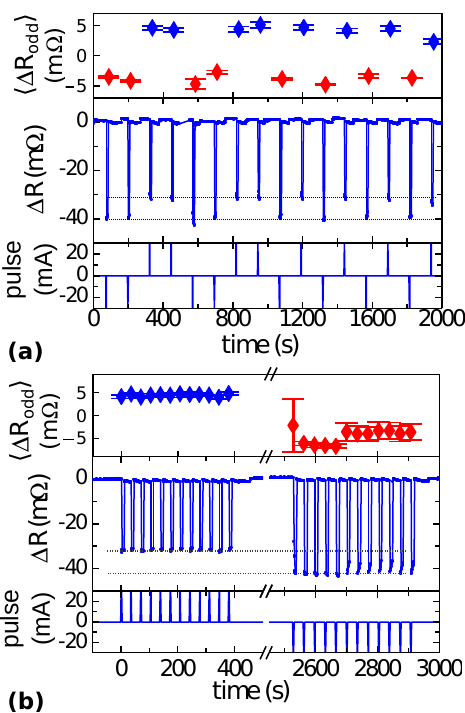}
\caption{Resistance change of a 300~nm diameter nanopillar in response to pulse trains applied while temperature is held at 404~K on the heating branch. (a) Response to alternate negative and positive pulse trains, showing the timing of the current pulse trains, the time series of $\Delta R$ measurements, and $\langle \Delta R_\mathrm{odd} \rangle$ during each resistance excursion. (b) Response to repeated pulses of the same sign, with the same quantities plotted. In each case, dotted lines indicate the difference in response owing to current sign. All pulse trains shown consist of $10^6$ pulses, each of 1~$\mu$s duration, with an amplitude of 20~mA, delivered evenly over a 1~s period. (The pulse sequence in (b) was taken after cooling and heating the sample back to 404~K after measuring (a) giving a good demonstration of the reproducibility of the effect).}
\label{pulsemem}
\end{figure}

The third and final difference is that the sign of the current pulse now matters. Whilst the resistance excursions $\Delta R$ are all negative in this case, the amplitude depends on the direction of current flow during the pulse train. For positive current flow $\Delta R \approx -40$~m$\Omega$ whilst for negative current flow $\Delta R \approx -32$~m$\Omega$. This difference appears consistently, regardless of whether the current direction is intermittently switched (Fig.~\ref{pulsemem}a) or the pulse trains are repeatedly applied in the same direction (Fig.~\ref{pulsemem}b).

We thus decompose $\Delta R$ into two contributions. There is a contribution $\Delta R_\mathrm{even} \approx -36$~m$\Omega$ that is even in current that we can attribute to Joule heating. The jagged and hysteretic nature of the resistance-temperature curve in the mixed phase regime makes it difficult to reliably determine what temperature change this corresponds to in a direct way, but given that it is the same pulse train through the same nanopillar it is likely to be similar to the 1.4~K rise in the control experiments.

The other contribution is odd in current and has magnitude $\Delta R_\mathrm{odd} \approx \pm 4$~m$\Omega$, with the sign given by the sign of the corresponding conventional current. Average values $\langle \Delta R_\mathrm{odd} \rangle$ during each resistance excursion are shown in the top parts of Fig.~\ref{pulsemem}(a) and (b). A rise in resistance corresponds to a greater AF phase fraction, meaning that the phase boundary has moved down, in the direction of the electron flow. Conversely, a drop in resistance, indicating a larger FM phase fraction, can be associated with the phase boundary moving up in the direction of the electron flow. The resistance changes are thus consistent with the phase boundary moving in response to the injection of spin-polarised electrons from the FM into the AF phase material, or unpolarised electrons being driven from the AF material into the FM phase region \cite{Zyuzin2010}. This current-induced change in resistance is a memristance\cite{Chua1971}, in this case of magnitude 4~m$\Omega$, in response to a total charge of 20 mC of charge passing through the device. It is nevertheless transient, since the device resistance always relaxes to the same level after the pulse, showing that the phase boundary returns to its original pinning site. Indeed, the $R(T)$ curve shown in Fig.~\ref{transition} shows no sharp steps around this point, so this is not unexpected.

From the data in Fig.~\ref{transition} we can see that the total change in resistance from a fully AF to fully FM state is $\sim -30$~m$\Omega$. Since there is 60~nm of FeRh magnetic phase change alloy in the film from which the pillar was patterned, our measured value of $\Delta R_\mathrm{odd} \approx \pm 4$~m$\Omega$ corresponds to the phase boundary being reversibly displaced by $\sim \pm 8$~nm.

We can compare our observations to the predictions of spin accumulation theory \cite{Zyuzin2010} by calculating the effective field exerted on the AF material by spins injected from the FM at a distance $x$ from the phase boundary according to
\begin{equation}
B_\mathrm{eff}(x) = \frac{eJ}{g \mu_\mathrm{B}} \frac{2 R_\mathrm{FM} R_\mathrm{AF}}{R_\mathrm{FM} + R_\mathrm{AF}} P \exp \left( -x/\ell_\mathrm{FM} \right), \label{eff_field}
\end{equation}
where $e$ is the electron charge, $J$ is the current density, $g=2$ is the Land\'{e} g-factor, and $\mu_\mathrm{B}$ is the Bohr magneton. The resistance-area products are given by
\begin{equation}
R_\mathrm{FM} = \ell_\mathrm{FM}\frac{\sigma_\uparrow + \sigma_\downarrow}{4 \sigma_\uparrow \sigma_\downarrow}, \quad R_\mathrm{AF} = \ell_\mathrm{AF} \frac{1}{\sigma_\mathrm{AF}},
\end{equation}
in which $\sigma_{\uparrow,\downarrow}$ are the spin-resolved conductivities in the FM phase, $\sigma_\mathrm{AF}$ is the conductivity of the AF phase, and $\ell_\mathrm{FM,AF}$ are the spin diffusion lengths in the FM and AF phases, respectively. The spin polarisation of the current in the FM phase is given by the usual expression
\begin{equation}
P = \frac{\sigma_\uparrow - \sigma_\downarrow}{\sigma_\uparrow + \sigma_\downarrow}.
\end{equation}
Directly determining the resistivity $\rho$ of our pillar in the two phases is complicated by the fact that the measured resistance is dominated by lead and contact resistances in our quasi-four point geometry. We can refer to our previous resistivity measurements of FeRh thin films \cite{DeVries2013} to estimate the resistivities in the relevant temperature range as being $\rho_\mathrm{AF} \approx 170$~$\mu \Omega$cm and $\rho_\mathrm{AF} \approx 90$~$\mu \Omega$cm. We used a typical ferromagetic transition metal value \cite{Bass1999,campbellfert1982} of $P \approx 0.75$ in the diffusive regime to estimate the spin resolved conductivities in the FM phase.

There is little extant data on the spin diffusion length in FeRh. Here we use the same estimate as Suzuki et al., that $\ell_\mathrm{AF} \approx 0.5$~nm \cite{Suzuki2015} and use an Elliot-Yafet scaling $\ell \sim \rho^{-1}$ (not always very well obeyed \cite{Bass2007}), giving $\ell_\mathrm{FM} \approx 0.9$~nm.

At our injected current density $J = 2 \times 10^{11}$~A/m$^2$, Eq.~\ref{eff_field} yields a value of $\sim 1$~T close ($x$ a few lattice constants) to the phase boundary. Such a field would reduce the transition temperature by $\sim 8$~K in a typical FeRh sample \cite{Maat2005,Massey2020}. Comparing the temperature range needed to complete the transition in Fig.~\ref{pulsemem} ($\sim 50$~K) with the FeRh film thickness from which the pillar was patterned ($50$~nm) yields a transition temperature gradient of $\sim 1$~K/nm, and so a suppression of 8~K would correspond to a spin injection-driven phase boundary displacement of 8~nm, in good agreement with the experimental observation.

\begin{figure}[tb]
\includegraphics[width=7.5cm]{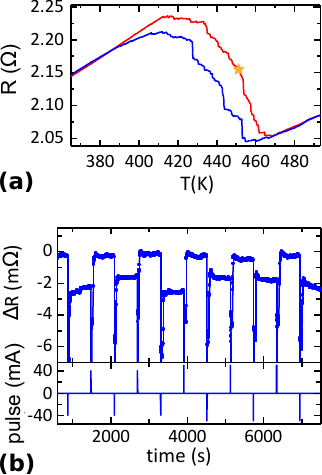}
\caption{Hysteretic response to pulsed currents. (a) Resistance with temperature for a a 500~nm diameter nanopillar. The star at 452~K on the heating branch, close to the transition mid-point, gives the point at which (b) the resistance changes due to pulsed currents were measured. The pulse trains again consist of $10^6$ 1~$\upmu$s pulses at a 10\% duty cycle, but in this case the current amplitude was 40~mA for the first five pulse trains and 50~mA for the last six. The resistance of the nanopillar is reversibly changed between two long-lived values by pulse trains of opposite current polarity.} \label{pulsememhyst}
\end{figure}

We show data acquired from a 500~nm diameter pillar patterned from the same film in Fig.~\ref{pulsememhyst}. In this case the temperature was held at 452~K, close to the midpoint of the transition, and so the phase boundary will begin in a pinning site approximately halfway up the nanopillar. The $R(T)$ curve for this device shows sharp steps in this region, indicating the presence of multiple pinning sites for the phase boundary. Pulse trains with the same time structure but of 40 or 50~mA amplitude were applied. The corresponding current density is $\sim 2 \times 10^{11}$~A/m$^2$. There is an initial negative jump in $\Delta R$ due to heating caused by each pulse train that is on the scale several tens of m$\Omega$. The plot is truncated to show that this transient excursion in $\Delta R$ is followed by cooling back to one of two long-lived resistance states, separated by about 2~m$\Omega$. Which state is returned to depends on the polarity of the current pulse train, containing 40 or 50~mC of charge, that was previously applied in a manner that is again consistent with the phase boundary having moved in the direction of electron flow. This long-lived behaviour is likely to owe to the availability of different pinning sites between which the phase boundary can be moved at this point in the transition, and demonstrates the memory aspect of the memristive response.

To summarise, we have patterned thin film FeRh into nanopillars capable of carrying a vertical current flow. By using a doping growth gradient we were able to guarantee a phase domain boundary wall within the nanopillar perpendicular to that current flow. Passing current pulses through the nanopillar has been shown to move this phase boundary repeatably and reversibly. The measured resistance signal can be split into two parts. First is the Joule heating effect, which is even in current direction. Second is a motion dependence on current direction giving rise to a few m$\Omega$ change in junction resistance that is odd in current. The size of signal we observe is consistent with theoretical predictions and so we attribute this signal to spin injection through the FM/AF phase boundary. This $\Delta R$ can be either transient or long-lived depending on the details of the phase boundary energy landscape, and represents a memristance arising from its motion. Our doping gradient leads to a flat phase boundary \cite{LeGraet2015}, circumventing the scaling issues often found in oxide memristors where current hotspots form due to the presence of conducting filaments \cite{Jo2009}. Scaling our nanopillars to smaller diameters will increase $\Delta R$ whilst reducing the current needed to achieve a given current density. This will reduce the total charge needed to achieve a given resistance change, improving the memristive properties. Bespoke doping profiles will allow the operating point and dynamic range of our devices to be engineered to suit different applications. Moreover, our device design involves only a single nanopillar, requiring many fewer lithography squares than spin memristor realisations based on lateral magnetic domain wall motion \cite{Muenchenberger2012,Lequeux2016,Fukami2016} or the collective response of large numbers of pillars \cite{Raymenants2018}, and does not require exotic fabrication methods \cite{Al-Bustami2018}. These considerations ease its potential adoption in neuromorphic spintronic circuits\cite{Jo2010}.

\begin{acknowledgments}
This work was supported by EPSRC grants EP/M018504/1 and EP/M019020/1 and by the Diamond Light Source.
\end{acknowledgments}

\section*{Data Availability}
Data associated with this publication is available from the University of Leeds repository at [DOI to  be confirmed].

\bibliography{library}


\end{document}


\title{Phase domain boundary motion and memristance in gradient-doped FeRh nanopillars induced by spin injection: Supplementary Information}

\author{Rowan~C.~Temple}
\affiliation{School of Physics and Astronomy, University of Leeds, Leeds LS2 9JT, UK}

\author{Mark~C.~Rosamond}
\affiliation{School of Electronic and Electrical Engineering, University of Leeds, Leeds LS2 9JT, UK}

\author{Jamie~R.~Massey}
\affiliation{School of Physics and Astronomy, University of Leeds, Leeds LS2 9JT, UK}

\author{Trevor~P.~Almeida}
\affiliation{SUPA, School of Physics and Astronomy, University of Glasgow, Glasgow G12 8QQ, UK}

\author{Edmund~H.~Linfield}
\affiliation{School of Electronic and Electrical Engineering, University of Leeds, Leeds LS2 9JT, UK}

\author{Damien~McGrouther}
\affiliation{SUPA, School of Physics and Astronomy, University of Glasgow, Glasgow G12 8QQ, UK}

\author{Stephen~McVitie}
\affiliation{SUPA, School of Physics and Astronomy, University of Glasgow, Glasgow G12 8QQ, UK}

\author{Thomas~A.~Moore}
\affiliation{School of Physics and Astronomy, University of Leeds, Leeds LS2 9JT, UK}

\author{Christopher~H.~Marrows}
\email{c.h.marrows@leeds.ac.uk}
\affiliation{School of Physics and Astronomy, University of Leeds, Leeds LS2 9JT, UK}

\date{\today}

\maketitle

\section{Methods}

In this section we give further details of the experimental methods used in this work.

\subsection{Growth and characterisation}

The doped FeRh film was grown by DC sputtering at 600$\degree$C onto commercially obtained MgO (001) substrates\cite{LeGraet2013}. The stack sequence is MgO/NiAl(70nm)/FeRh(Pd,Ir)(50nm) where the NiAl layer is an epitaxially matched metal used for the bottom contact to the nanopillar \cite{Kande2011}. The doped FeRh layer was sputtered from a Pd and an Ir doped FeRh target. Angled magnetrons were used to allow rapid switching between target material growth and appropriate Pd or Ir densities were achieved by multiple 1~nm layer growths, each layer consisting of the appropriate ratio of Pd to Ir doping. A final anneal at 700$\degree$C for 1~hr in vacuum allowed the FeRh to obtain the proper crystal structure and smooth out the doping profile within the film thickness. The film examined here is NiAl(70nm)/ Fe$_{50}$Rh$_{47.2}$Pd$_{2.8}$(25nm)/ Fe$_{50}$Rh$_{47.1}$Pd$_{2.2}$Ir$_{0.7}$(15nm)/ Fe$_{50}$Rh$_{46.8}$Pd$_{1.7}$Ir$_{1.5}$(10nm).

\subsection{Patterning}

The bottom electrodes and nanopillars were defined using e-beam lithography on a JEOL JBX-6300FS system. Ar ion milling through a ma-N2403 resist mask was used for pattern transfer \cite{Temple2018}. The flying bridge contacts were created in a PMMA/MMA based lift-off process. An electron dose profile across the bridge used the contrast shift between the PMMA and MMA layers to define contact and flyover regions. The bridges contacts were sputter deposited Ti(6nm)/Au(300nm). An insert PMGI resist layer was used to create undercut so as to improve lift-off.

\subsection{Measurements}

The HAADF and EDS imaging and analysis shown in Fig.~\ref{nanopillar}c of the main text were carried out on JEOL Atomic Resolution Microscope (JEM-ARM200F) TEM, operating at 200~kV. The elemental data was acquired with a Bruker XFlash EDS detector.

Transport measurements were performed in a liquid nitrogen cooled Oxford Instruments OptistatDN with a temperature range of 77-500~K. A Keithley 6221 current source was used for pulsed current injection and signals were detected using a Keithley 2182 nanovoltmeter in the quasi-four point geometry shown in Fig.~\ref{nanopillar}b of the main text. All measurements were carried out at zero applied field.

\section{Current-driven heating}

\begin{figure}[tb]
\includegraphics[width=7.5cm]{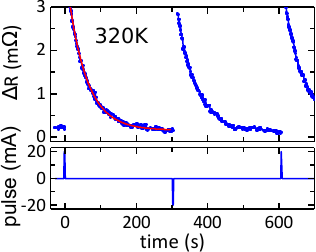}
\caption{Heating due to current pulses. The top panel shows resistance change response to current pulse trains through the 300~nm diameter nanopillar at 320~K. The pulse train times and current amplitude are shown in the bottom panel. Each pulse train consists of $10^6$ individual pulses each lasting 1~$\mu$s, delivered with a 10\% duty cycle. This allows a 9~$\mu$s cooling period after each pulse, which proved necessary to increase the total charge delivered while managing the heating effects of the current. The pulse response is symmetric with positive and negative currents as expected for Joule heating effects. The red line shows an exponential decay fit to the post-pulse cooling. The fitting parameters give an equivalent heating amplitude of 1.4~K and a decay time of 54~s.}
\label{pulseheat}
\end{figure}

A side-effect of high current densities in nanoscale devices is Joule heating, which will affect the device resistance. In Fig.~\ref{pulseheat} here, we show the effect of applying a pulse train of $10^6$ pulses, each of 1~$\upmu$s duration, with a duty cycle of 10\% to the 300~nm nanopillar device. The amplitude of each pulse was 20~mA, corresponding to a current density of $\sim 3 \times 10^{11}$~A/m$^2$. This experiment was carried out at 320~K, a temperature at which the nanopillar is fully AF and no phase boundary is present, as can be seen in Fig.~\ref{pulsemem} in the main text.

The measured resistance of the nanopillar rises by $\Delta R \approx 3$~m$\Omega$ at the time of the pulse before exponentially relaxing back to its equilibrium value on a $1/e$ timescale of 54~s. By making a comparison with the linear regime in Fig.~\ref{transition} of the main text, we determined that this resistance rise corresponds to a temperature rise of about 1.4~K. The response is the same regardless of the direction in which the current pulses flow, showing that in this case the resistance change is entirely due to Joule heating.

Test nanopillars were stable at this duty cycle, but often failed when the 9~$\upmu$s gap between pulses was reduced much below this value. This indicates that the nanopillar temperature briefly rises well above this value but cools on a $\upmu$s timescale, fast enough to be beyond the time resolution of these measurements. The principal heatsink for cooling will be the region of substrate just under the pillar, and we attribute this temperature rise and slow decay to the heating of that region and gradual cooling as heat is conducted away into the rest of the substrate.

\bibliography{library}